# Decoding Parkinsonian Tremor: An Explainable Framework Integrating Multi-Revolution Spatial and Spectral Dynamics of Spiral Drawings

**Tharaka Wijethunge1, Maheshi Dissanayake2*, and Sajitha Weerasinghe3**

[1]Department of Electrical, Computer, and Systems Engineering, Rensselaer Polytechnic Institute, Troy, New York, United States of America

[2]Department of Electrical and Electronic Engineering, Faculty of Engineering, University of Peradeniya, Peradeniya, Sri Lanka

[3]Neurology Unit, Teaching Hospital Peradeniya, Peradeniya, Sri Lanka

***Corresponding author:** Maheshi Dissanayake (maheshid@eng.pdn.ac.lk )

---

## Abstract

Parkinson's disease (PD) manifests in motor impairments that are detectable through digitized spiral drawings. This study introduces an explainable framework for PD screening using a novel radial-sampling feature fusion approach. We transform 2D spiral images into 1D revolution signals via a systematic ray-sampling technique to extract three distinct revolutions. We integrate spatial metrics, such as inter-revolution spacing variability and RMS radial derivatives, with spectral descriptors derived from Fast Fourier Transform (FFT) analysis across low, mid, and high harmonic bands. A total of 20 features were utilized to train state-of-the-art machine learning models, including Support Vector Machines (SVMs), Random Forests (RFs), and Light Gradient Boosting Machines (LightGBMs). Among these, the RF classifier demonstrated superior performance. Subsequent 5-fold cross-validation stability analysis along with feature importance analysis identified RMS radial derivative of the outer revolution ($r3_deriv_rms$) as the most critical biomarker. Stratified Cross-Validation demonstrates that combining spatial and frequency features significantly enhances detection accuracy compared to single-domain methods, facilitating effective clinical deployment even in data-scarce environments. This interpretable pipeline provides a robust, low-cost "white-box" screening tool, offering a practical alternative to opaque deep-learning models for early clinical intervention.

## 1. Introduction

Parkinson's disease (PD) is a progressive neurodegenerative disorder that impairs motor control, producing clinical signs such as tremor, rigidity, bradykinesia, postural instability, cognitive changes, mood disorders, sleep disturbances, and autonomic dysfunction [1]. Multiple studies used standardized handwriting and drawing tasks to quantify PD motor deficits such as bradykinesia [2-4]. The earliest symptoms are considered alterations in handwriting kinematics [5]. More broadly, handwriting abnormalities in PD extends beyond micrographia and include impairments in kinematic, pressure, and fluency-related parameters[6]. Prior work reports characteristic PD-related handwriting abnormalities, including reduced writing amplitude and impaired letter formation, as measured by standardized writing tasks [7,8], although Archimedean spiral tracing has been widely studied as a low-cost, non-invasive proxy for detecting this motor dysfunction [9,10]. At present, there isn't a definitive test for PD, and the rate of misdiagnosis, particularly by non-experts, tends to be quite significant, with the chance of an incorrect diagnosis potentially hitting 20%[11]. Digitized versions of Archimedes spirals provide a computer-based, objective way to quantify drawing impairment and support automated classification in Parkinson's disease [12-14]. Even brief, routine handwriting tasks have been shown to differentiate PD patients from healthy controls, supporting the use of handwriting as a practical, objective biomarker [15].

In recent years, deep learning has showcased impressive capabilities in the realm of image classification. Specifically, Convolutional Neural Networks (CNNs), a specific subset of deep learning models, have been notably effective in biomedical image classification tasks. While deep learning holds significant promise for image classification, it is essential to recognize that it requires considerable computational resources, large datasets, and specialized GPUs [16], for effective training. While deep CNNs can achieve high accuracy on large curated datasets [17], practical model trainings often operate with small-samples, with device heterogeneity, and require interpretable decisions that clinicians and patients can trust, rather than "black-box" type behaviors[18].

Current clinical assessment of PD largely relies on semi-quantitative scales, such as the Unified Parkinson's Disease Rating Scale (MDS-UPDRS), which are susceptible to significant inter-rater variability and subjective interpretation. Consequently, there is a pressing need for objective, digital biomarkers that can provide reproducible measures of motor dysfunction. The limitations in the practical applicability of existing automated approaches—often characterized by high data dependency or 'black-box' architectures—motivated the development of a simpler, more explainable framework for identifying PD through spiral drawings.

This study addresses the gaps in literature by providing an interpretable pipeline that translates complex spiral trajectories into clinically meaningful spatial and spectral indices. Rather than applying a ‘black box’ type model, such as popular CNN models, directly to raw images, our method evaluates the spiral through a sequence of image analysis and feature extraction steps designed to handle diverse patient drawings while maintaining structural validity. We integrate multi-revolution spatial features with detailed spectral dynamics to capture subtle motor irregularities, such as tremor-induced fluctuations and inter-revolution spacing variability.

The proposed framework provides substantial clinical significance by offering an objective tool for characterizing Parkinsonian tremor through a non-invasive, low-cost approach. It should be noted that literature supports and presents evidence for the clinical relevance of the digitized Archimedes spiral test while highlighting the importance of methodological standardization [19]. The explainable architecture proposed in this research enhances clinician trust by clearly validating the biomechanical and frequency-based features influencing the model’s decisions, aligning with the growing demand for transparent digital biomarkers in neurology. Because the approach is easy to administer, it can be seamlessly incorporated into routine clinical workflows, telemedicine, and home-based monitoring. This work enables reliable longitudinal tracking of disease progression and quantification of therapeutic response—specifically for medication adjustments or deep brain stimulation—ultimately supporting individualized, data-driven management of Parkinson’s disease.

## 2. Related Work

The majority of literature in this domain primarily focuses on training large-scale deep learning (DL) models to identify PD. Specifically, transfer learning using architectures such as DenseNet and ResNet50 has demonstrated high performance; however, these models typically require an extensive number of spiral images for training [1]. Another approach, proposed by Li et al., focuses on combining genetic data and structural MRI for image acquisition. While effective, this method is not cost-effective and poses significant challenges for wider deployment [17]. In addition to transfer learning techniques, the literature indicates that Grad-CAM has been utilized to provide model explainability. Nevertheless, similar to other transfer learning applications, the volume of images required for training remains extensive [20].

In the handcrafted feature analysis approach, movement-error quantification techniques calculate discrepancies based on the distance between the template spiral and the actual drawing. These methods emphasize strong point correspondence, such as equivalent-angle pairing. While these approaches are interpretable, they mainly concentrate on geometric errors and do not explicitly account for physiological tremor signatures in the frequency domain [21,9]. Although smartphone deployment in practical settings has been covered in some publications, their focus was mainly on tremor and may be limited in robustness [22,23].

Moreover, several studies have enhanced the reliability and interpretability of spiral metrics by tackling practical measurement challenges, such as center detection and improved spiral utility [24]. Additionally, some researchers have investigated specific indicators, like temporal irregularity in spiral drawings [25]. Others have focused on validating graphical and drawing tasks to aim for clinically meaningful outcomes rather than simply categorizing results [3,26]. In addition, kinematic and pressure-based handwriting variables have been reported as useful for the differential diagnosis of PD[27]. These feature-driven approaches can be sensitive to the tasks performed and the preprocessing methods used, and it is important to note that effective features may not transfer well across different devices or populations without following a full end-to-end framework. Prior handwriting-based decision-support studies have similarly shown that carefully engineered markers can support PD discrimination without relying exclusively on end-to-end deep architectures [28]. A broader feature taxonomy combining kinematic, geometrical, and nonlinear handwriting descriptors has also shown strong relevance for modeling PD-related motor deficits[29].

Numerous studies support the use of spiral analysis as an objective indicator of motor impairment in PD. Research highlights that these measures provide valuable diagnostic information even in the early stages of the disease and are significantly correlated with changes in upper-limb function and pharmacological response [30,26]. Furthermore, evidence suggests that spiral irregularities may be influenced by factors beyond disease severity, such as medication-related complications; this underscores the necessity of accounting for clinical status when interpreting spiral-based metrics [4,26].

Recent advancements have enhanced the accessibility of PD spiral data collection through graphic table-based interfaces and smart pens, facilitating low-cost screening outside of specialized laboratory settings [31,32]. While various learning-based approaches currently classify PD by extracting frequency-domain features from digitized patterns and temporal data, emerging research is shifting toward real-time, tablet-based AI studies to evaluate feasibility in clinical environments [31,33-36]. Consequently, there is a critical need for small-scale inference and feature-processing, machine learning (ML) frameworks capable of operating on edge devices. This need is reinforced by studies showing that biometric handwriting analysis can support not only PD detection but also disease grading in clinically meaningful settings[37]. Developing systems that can retrain on new samples with limited computational resources while providing model explainability offers a more robust and practical alternative to traditional, resource-intensive deep learning methods. Recent studies in medical image classification have further highlighted the value of hierarchical feature fusion, ensemble learning, and attention-enhanced CNN designs for improving robustness and generalization across heterogeneous biomedical datasets [38,39].

## 3. Data and methods

### 3.1. Dataset

We used two sources of spiral drawings: (i) a set collected in a clinical setting using a simple, fixed procedure, and (ii) a small, carefully selected group from public websites to include different types of paper, pens, and lighting [40,41]. The clinical data of the study was collected from the teaching hospital of the Central province which has a large geographical catchment area. All the patients were assessed and diagnosed by a single neurology team so the interobserver variants would be minimal. To ensure diagnostic specificity, all participants underwent differential screening to exclude essential tremor, dystonic tremors, and metabolic-induced tremors, focusing exclusively on idiopathic PD as per Movement Disorder Society- Unified Parkinson's Disease Rating Scale (MDS-UPDRS) criteria. All the patients were given A4 size white-blank paper and a standard ball point pen. The patients were asked to sit comfortably in the clinic with their arm resting on the table. Next, they were asked to study the given reference spiral image and draw a simillar spiral in the blank sheet of paper without lifting the hand off the paper. Figure 1 displays typical sample drawings of a healthy person and that of a person with severe Parkinson's disease.

These drawings were either scanned or photographed and subsequently processed using a consistent set of steps for each image: straightening the page, adjusting the lighting and background, converting to grayscale, binarizing to emphasize the ink, and tightly cropping around the spiral. We resized the images to a uniform square dimension while preserving their original shape and aspect ratio. We removed any drawings that were incomplete, had heavy shadows, or displayed other obvious issues. For the training and test set, we retained images with clear strokes and sufficient contrast so that the full spiral is visible. This approach enables us to evaluate how effectively our methods work on everyday photos, rather than just on clinic scans. Moreover, if a subject had multiple spirals, we selected at most one for model training or testing to prevent overlap. When possible, we included simple notes to help us assess performance across different groups. Following this careful screening 30 spirals were collected at clinical setting by the authors.

To enhance model generalizability, we combined our clinical dataset with spiral images from previously published, publicly available PD benchmark datasets [40,41] to create the final training dataset used for model development. Furthermore, a healthy control cohort was incorporated to serve as a baseline reference for comparative analysis. The final consolidated dataset comprised 116 high-resolution spiral images, providing a diverse foundation of both symptomatic and healthy signatures for robust feature extraction and classification. This balanced approach ensures that the model can effectively distinguish subtle motor impairments from normal physiological variations. We initially developed and optimized our models using public image datasets [40,41] and subsequently refined the model with clinical data. Unless otherwise specified, we use a 70/15/15 train/validation/test split, ensuring class balance and, when feasible, maintaining severity levels.

Additionally, we conduct cross-source tests, where models trained on public images are directly evaluated on a clinical test set that has not been previously accessed.

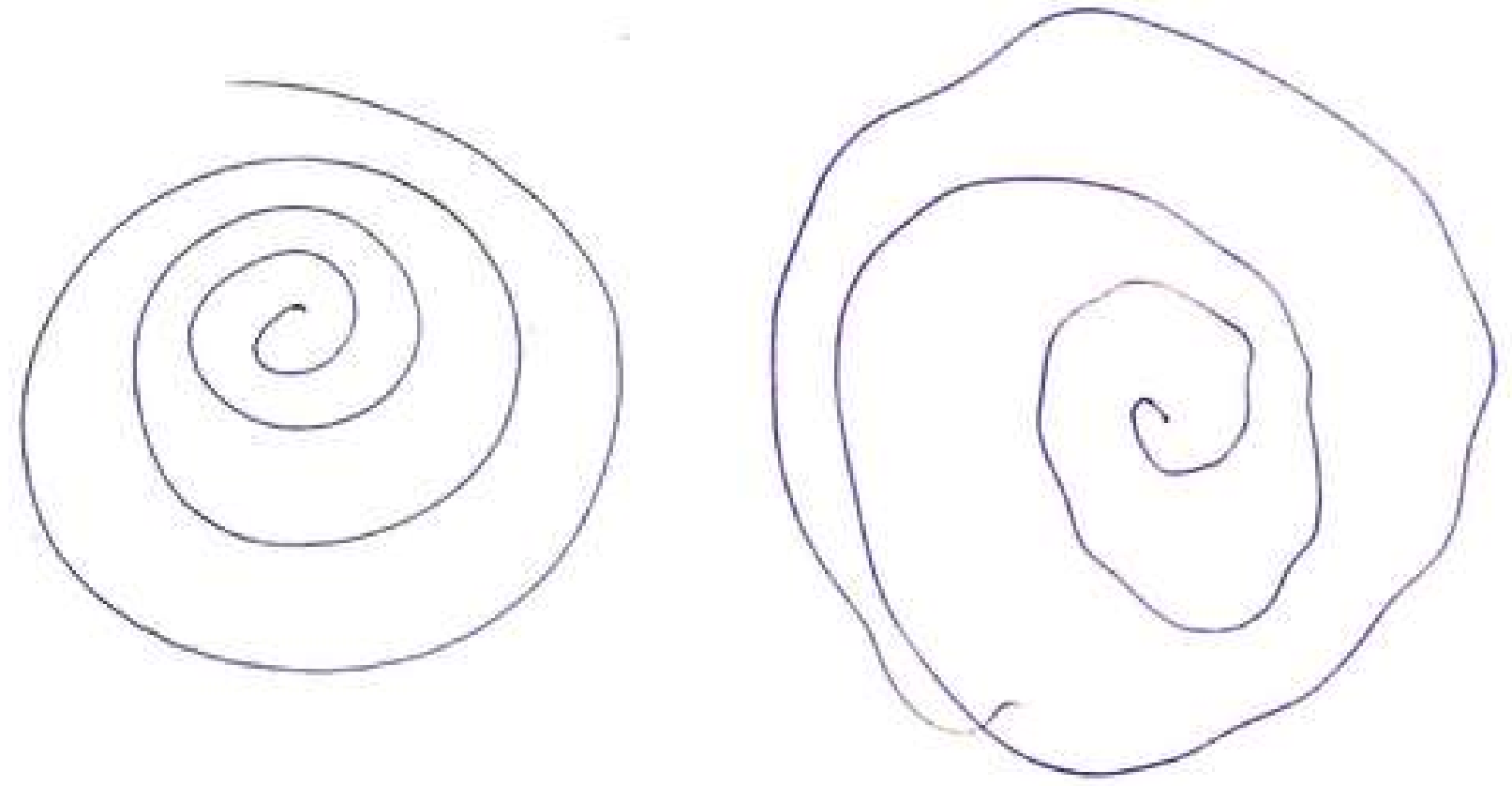

*Figure 1. Healthy Spiral Drawing (left) PD Spiral Drawing (right)*

### 3.2. Radial Distance Extraction From Spiral Drawing

#### 3.2.1. Spiral Segmentation

Each RGB spiral image, $I \in R^{H\times W\times 3}$, is first converted to a grayscale representation, $G$, where each pixel denotes a normalized intensity value. To enhance the signal-to-noise ratio, $G$ is pre-processed using either a median or bilateral filter for noise reduction. Subsequently, a binary mask $B \in [0,1]^{H\times W}$ is generated via global thresholding according to Eq. (I):

$$B(i,j) = \begin{cases} 1, & G(i,j) \leq \tau^* - \text{ (i.e. belongs to foreground)}, \\ 0, & otherwise - \text{ (i.e. belongs to background)}. \end{cases} \quad \text{(I)}$$

where $\tau^\star$ represents the optimal threshold selected to maximize the separability between the foreground (ink) and background classes. To ensure robustness against non-uniform illumination and scanning artifacts, we incorporate adaptive thresholding based on local brightness variations. This is followed by morphological opening and closing operations using a *3x3* structuring element to eliminate isolated artifacts and bridge small discontinuities in the binarized trace, thereby ensuring a continuous and representative spiral path.

#### 3.2.2. Spiral Center Detection using Circular Statistics

To accurately characterize the spiral geometry, the center, $\hat{c} = \left(\widehat{c_x}, \widehat{c_y}\right)$, must be precisely identified. Let $S$ denote the set of foreground pixel coordinates $(x_k, y_k)$ extracted from the binary mask $B$ such that P = {(x, y) ∣ B(x, y) = 1}. We estimate the optimal spiral center by identifying the candidate point $c$ for which the foreground pixels exhibit the most uniform angular coverage when expressed in polar coordinates. Unlike traditional centroid-based methods, this angular-uniformity approach provides robustness against image artifacts, scanning noise, and non-uniform stroke distributions that typically bias the center of mass. Its effectiveness was empirically evaluated against centroid and distance-transform-based alternatives. Following the principles of circular statistics, the estimation is formulated as in Eqs. (II) – (IV).

$$\hat{c} = \arg\max_{c} V(c) \qquad \text{(II)}$$

$$V(c) = 1 - \left|\frac{1}{N_s}\sum_{k=1}^{N_s} e^{j\theta_k(c)}\right| \qquad \text{(III)}$$

$$\theta_k(c) = \angle\left((x_k, y_k) - c\right) \qquad \text{(IV)}$$

In this framework, $\theta_k(c)$ represents the polar direction of the $k$-th foreground pixel relative to a candidate center $c$, where $N_s$ is the total number of extracted foreground pixels ($N_s = |P|$). The term within the absolute value bars in Eq. (III) represents the mean resultant vector length; a value approaching zero indicates that the pixel directions are distributed uniformly around the center, thereby maximizing the circular variance $V(c)$. Consequently, the candidate coordinate that maximizes $V(c)$ is selected as the most geometrically centered origin, providing a stable basis for subsequent feature extraction.

### 3.2.3. Coordinate Transformation and Scale Normalization

To ensure that the extracted features are comparable across different drawings and varying acquisition scales, the raw spiral coordinates $(x_k, y_k)$ are transformed relative to the estimated center $\hat{c}$ and subsequently normalized by the maximum observed radius $R_{\max}$. This process maps the entire spiral trace to a unit circle, as defined in Eq. (V) and Eq. (VI):

$$\overline{x_k} = \frac{x_k - \widehat{c_x}}{R_{max}}, \qquad \overline{y_k} = \frac{y_k - \widehat{c_y}}{R_{max}} \qquad \text{(V)}$$

$$R_{max} = \max(k)\sqrt{(x_k - \widehat{c_x})^2 + \left(y_k - \widehat{c_y}\right)^2} \qquad \text{(VI)}$$

Following this transformation, the spiral's boundary maps to a radius of 1. All subsequent downstream computations, including spectral analysis and irregularity measurements, utilize these normalized coordinates $(\overline{x_k}, \overline{y_k})$ to mitigate the influence of drawing size and focus strictly on the morphological and kinematic patterns of the patient's handwriting.

### 3.3. Feature Extraction and Signal Preprocessing

#### 3.3.1. Radial Feature Extraction from Spiral Revolutions

To capture the fine-grained geometric properties of the hand-drawn spirals, we convert the two-dimensional binarized images into three distinct one-dimensional "revolution" (rev) signals through a systematic ray-sampling technique. This process initiates from the estimated center $\left(\widehat{c_x}, \widehat{c_y}\right)$ and utilizes a uniform angular grid, where each sampling angle $\theta_j$ is defined as in Eq. (VII).

$$\theta_j = j\Delta\theta, \quad j = 0, \dots, N-1, \quad \Delta\theta = 5^\circ, \quad N = \frac{360^\circ}{\Delta\theta} \qquad \text{(VII)}$$

For each angular step, a radial ray is cast from the center. We detect consecutive foreground (ink) pixels along this ray at small radial increments, identifying the first three valid intersections that satisfy a minimum revolution length and are separated by a distinct background gap. These intersections correspond to the first (innermost), second, and third spiral revolutions. Figures 2 and 3 illustrate sample labeling of Rev 1, 2 and 3, distinguished by three different colors to visualize the extraction process. The radius for each revolution, $\mathrm{r_k[j]}$, is recorded as the mean radial distance of all pixels within that specific intersection. This produces three discrete angular signals as in Eq. (VIII).

$$r_k[j] = r_k\left(\theta_j\right), \quad k \in \{1,2,3\}, \qquad \text{(VIII)}$$

where $r_k[j]$ is measured in pixels. In instances where a ray intersection is not identified at a specific angle, the value is set to $NaN$ to preserve the temporal integrity of the signal for downstream analysis.

### 3.3.2. Signal Conditioning and Missing Data Handling

Extracted signals $r_k[j]$ often contain discontinuities due to intermittent pen contact or faint strokes. To ensure signal integrity for analysis, we implement a two-stage circular restoration process: (i) Circular Linear Interpolation: Small angular gaps are filled via linear interpolation along the circular axis to maintain the periodic nature of the spiral trace. (ii) Circular Median Filtering: A median filter with window length $W$ and wrap-around indexing is applied to mitigate high-frequency noise while preserving geometric trends. While large gaps are excluded from general statistical analyses, they are restored via circular interpolation specifically for Fast Fourier Transform (FFT) computations to ensure the signal continuity required for spectral processing.

### 3.3.3. Residuals and Inter-revolution Spacing

To isolate local irregularities, we derive two additional 1D signal sets from the rev radii. First, each revolution is centered by its median to obtain the residual signal $e_k[j]$ as in Eq. (IX), which captures deviations from the typical radius while remaining robust to outliers:

$$e_k[j] = r_k[j] - \text{median}\{r_k[j]\} \qquad \text{(IX)}$$

Second, we quantify the consistency of the spiral's expansion by measuring the spacing $s_{ab}[j]$ between adjacent revolutions $(a, b) \in \{(1,2), (2,3)\}$ as in Eq. (X)

$$s_{ab}[j] = r_b[j] - r_a[j] \qquad \text{(X)}$$

These signals summarize how the gap between successive turns fluctuates with the angle, providing critical indicators of motor dysfunction.

## 3.4. Spatial Feature Extraction

Spatial features are derived from the raw revolution signals $r_k[j]$, their robust residuals $e_k[j]$, and the inter-revolution spacing $s_{ab}[j]$ as defined in Eqs. (VIII), (IX) and (X) respectively. All statistics are calculated over the subset of angles where the required signal values are finite.

### 3.4.1. Kinematic and Roughness Descriptors

These features quantify the high-frequency fluctuations and micro-irregularities characteristic of Parkinsonian tremors.

- **RMS Radial Derivative ($\boldsymbol{r}_{\mathbf{3}_\text{rms_deriv}}$):** Measures rapid angular fluctuations in the outer revolution by computing the Root Mean Square (RMS) of the residual derivative as in Eq. (XI).

$$\text{RMS}_{deriv} = \sqrt{\frac{1}{N}\sum_j (e_3[j] - e_3[j-1])^2} \quad \text{(XI)}$$

- **Zero-Crossing Rate ($\boldsymbol{r}_{\mathbf{1}_\text{zcr}}$):** Quantifies the frequency at which small angular fluctuations in the innermost revolution switch direction. Let $\mathbb{1}(\cdot)$ be the indicator function; the $\boldsymbol{r}_{\mathbf{1}_\text{zcr}}$ rate is given by Eq. (XII).

$$ZCR = \frac{1}{N-1}\sum_{j=1}^{N-1} \mathbb{1}(e_1[j] \cdot e_1[j-1] < 0) \quad \text{(XII)}$$

  A higher ZCR suggests increased oscillatory, tremor-like variations.

- **Robust Roughness ($\boldsymbol{r}_{\mathbf{2}_\text{res_mad}}$):** A measure of instability for the middle revolution based on the Median Absolute Deviation (MAD) of the residuals is given by Eq. (XIII).

$$MAD = \text{median}(|e_2[j] - \text{median}(e_2)|) \quad \text{(XIII)}$$

  Since $e_k$ is centered by its median, the second term is typically near zero, isolating the raw dispersion of the signal.

### 3.4.2. Geometric and Morphological Features

These metrics describe the global shape and continuity of the spiral.

- **Outer Radius Statistics ($\boldsymbol{r}_{\mathbf{3}_\text{mean}}, \boldsymbol{r}_{\mathbf{3}_\text{median}}$):** Summarize the typical scale of the spiral's outer boundary as in Eq. (XIV).

$$\mu_{r3} = \frac{1}{N}\sum r_3[j], \quad \widetilde{r_3} = \text{median}(r_3) \quad \text{(XIV)}$$

- **Max Missing Segment ($\boldsymbol{r}_{\mathbf{1}_\text{max_gap_deg}}$):** Identifies the largest contiguous angular gap in the first revolution. If $M$ is the maximum number of consecutive missing bins, the gap in degrees is given by Eq. (XV).

$$\text{Gap}_{deg} = M \cdot \Delta\theta \quad \text{(XV)}$$

### 3.4.3. Inter-revolution Consistency and Overlap

Inter-revolution features characterize the stability of the spiral's expansion.

- **Spacing Statistics ($s_{12_\text{mean}}, s_{12_\text{std}}, s_{12_\text{cv}}$):** For the gap between the first and second revolutions, $s_{12}[j]$, we compute the mean ($\mu_s$), standard deviation ($\sigma_s$), and coefficient of variation ($CV_s$) as in Eq. (XVI).

$$CV_s = \frac{\sigma_{s12}}{\mu_{s12}} \quad \text{(XVI)}$$

- **Revolution Crossing Fraction ($s_{12_\text{crossings}}$):** Measures the frequency of irregular overlaps (where the spiral "crosses" itself) by calculating the fraction of angles where spacing becomes negative as in Eq. (XVII).

$$\text{Crossing Rate} = \frac{1}{N}\sum_j \mathbb{1}(s_{12}[j] < 0) \quad \text{(XVII)}$$

- **Outer Spacing Consistency ($s_{23_\text{std}}$):** Captures the variability in the gap between the second and third revolutions to assess turn-to-turn uniformity.

## 3.5. Frequency-Domain Feature Extraction

Frequency-domain features used in this study are derived from the Discrete Fourier Transform (DFT) of each revolution signal. To ensure spectral continuity, missing samples in the radial signals are first restored via circular interpolation. Let $\tilde{r}_k[j]$ represent the interpolated version of $r_k[j]$. To isolate oscillatory content from the global spiral trend, the DC component is optionally removed following Eq. (XVIII).

$$\hat{r}_k[j] = \tilde{r}_k[j] - \text{mean}(\tilde{r}_k) \quad \text{(XVIII)}$$

The DFT magnitude is then computed for each harmonic index $m$ using Eq. (XIX).

$$M_k[m] = \left|\sum_{j=0}^{N-1} \hat{r}_k[j] e^{-j\frac{2\pi}{N}mj}\right| \quad \text{(XIX)}$$

Next, we categorize spectral energy into three harmonic bands to differentiate between geometric structure and pathological tremors: Low Band ($m = 2\ to\ 4$) which captures the fundamental spiral geometry, Mid Band ($m = 5\ to\ 8$) which represents intermediate fluctuations and High Band ($m\ =\ 9\ to\ 12$) that isolates high-frequency micro-tremors.

### 3.5.1. Spectral Band Energies

We quantify the spectral distribution by calculating the cumulative energy within specific harmonic bands (low, mid, and high), providing a profile of the tremor's intensity across different scales:

- **Band Energies ($E_{low}, E_{mid}, E_{high}$):** Computed as the sum of squared magnitudes within defined harmonic ranges as in Eq. (XX).

  $E_{band} = \sum_{m \in \text{band}} (M_k[m])^2$ (XX)

  For instant, the band energies for rev 1 (i.e. k=1) is represented as $E_{low} = \sum_{h=2}^{4} M_1[m]^2$, $E_{mid} = \sum_{h=5}^{8} M_1[m]^2$, and $E_{high} = \sum_{h=9}^{12} M_1[m]^2$. Simiallarly we can write for band energies for rev 2 and rev 3 following the Eq. (XX).

- **High-to-Low Spectral Ratio ($r_{1_\text{hilo_ratio}}$):** This ratio characterizes the dominance of high-frequency micro-tremors relative to lower-frequency geometric components as in Eq. (XXI).

  $Ratio_{HL} = \frac{E_{high}}{E_{low}}$ (XXI)

### 3.5.2. Dominant Oscillatory and Decay Components

To further distinguish pathological movements from healthy handwriting, we analyze the peak intensity and the rate of spectral roll-off:

- **Dominant Amplitude ($r_{1_\text{dom_amp}}$):** Captures the maximum non-DC magnitude, representing the most prominent oscillatory component in the first revolution as in Eq. (XXII).

  $A_{dom} = \max_{m>0} M_k[m]$ k ={1,2,3} (XXII)

- **Spectral Magnitude Decay ($r_{k_\text{spec_decay}}$):** Quantifies the rate at which spectral energy decreases as the harmonic index increases. This is determined by fitting a linear regression to the log-magnitude spectrum as in Eq. (XXIII).

  $$\log(M_k[m]) \approx \alpha \cdot m + \beta \qquad \text{(XXIII)}$$

  A more negative decay constant ($\alpha$) indicates a steeper drop-off, signifying a lack of high-frequency content typical of healthy, smooth handwriting.

### 3.5.3. Spectral Cross-Correlation

To assess the consistency of motor dysfunction across the drawing task, we measure the spectral similarity between the first and second revolutions using the Pearson correlation coefficient over shared harmonic indices as in Eq. (XXIV).

$$\rho_{12} = \text{corr}(M_1[m], M_2[m]) \qquad \text{(XXIV)}$$

### 3.6. Classification and Performance Evaluation

In this study, a total of 20 features—comprising the spatial and frequency-domain descriptors detailed in Sections 3.4 and 3.5—were utilized to train and evaluate popular state of the art machine learning models to detect presence of PD. The final preprocessed output, illustrating the clear morphological and spectral distinctions between a Healthy Control and a patient with PD, is depicted in Figures 2 and 3. These figures visualize the extraction process by highlighting three spiral revolutions (Rev 1, 2, and 3) in distinct colours. As shown in the provided sample images, the first (innermost), second, and third intersections with the radial ray are clearly identified. Furthermore, each individual sampled pixel position is marked to emphasize the differences in stroke consistency and geometric regularity between the two groups.

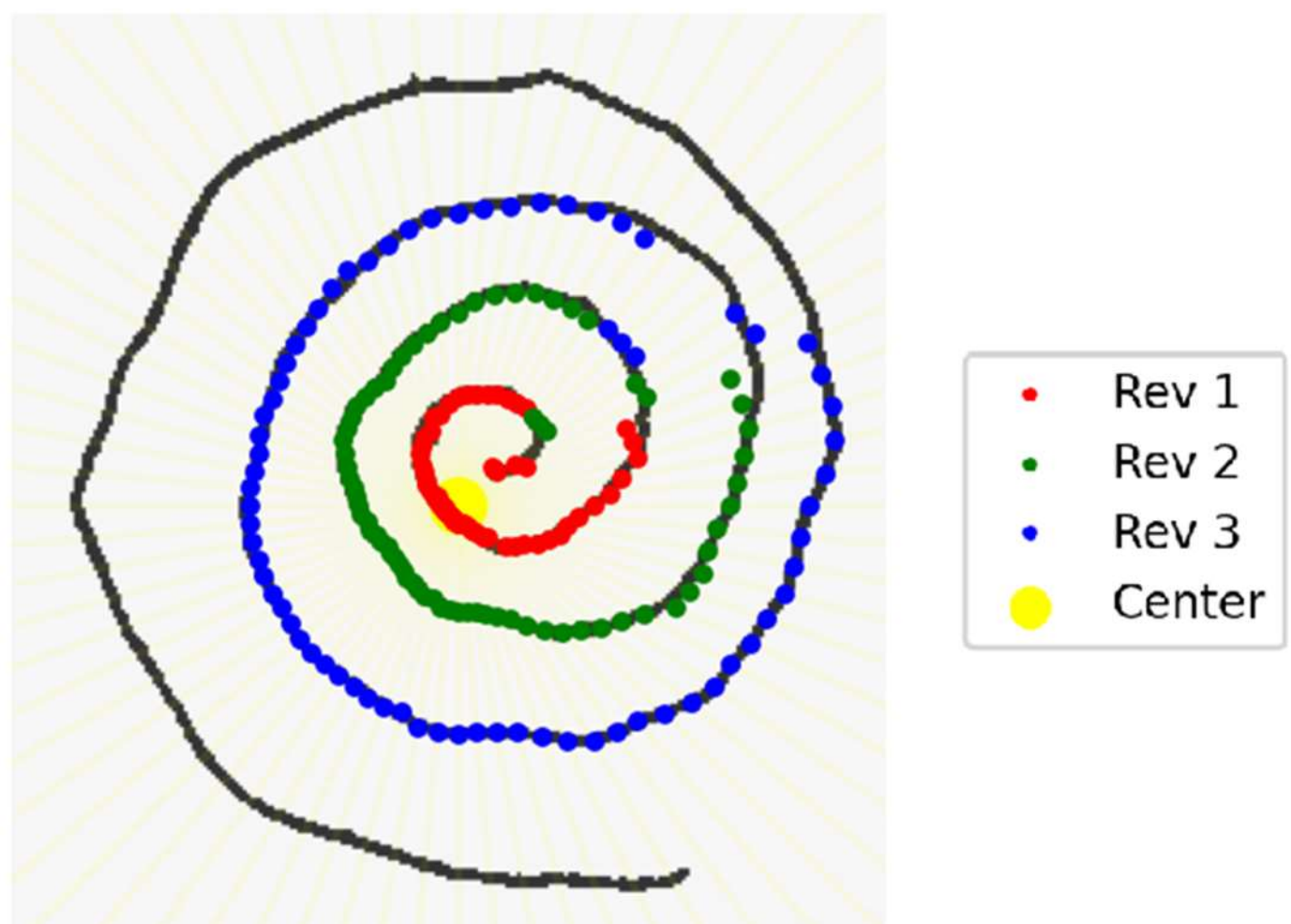

*Figure 2. Fully Preprocessed output used for feature extraction for the Healthy sample*

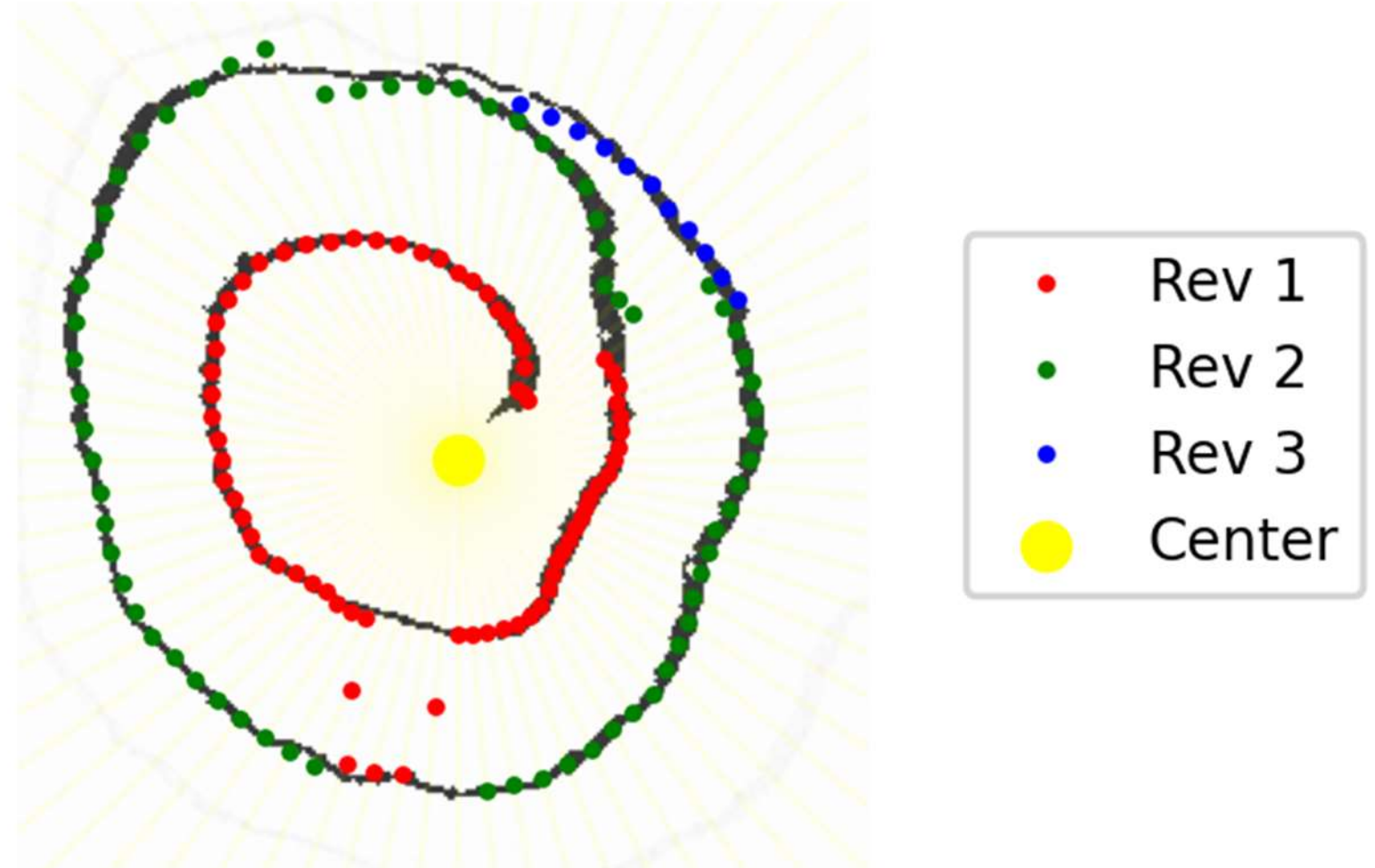

*Figure 3. Fully Preprocessed output used for feature extraction for the PD sample*

## 4. Machine Learning Model Training

We treat PD screening from spiral drawings as a supervised binary classification problem using the extracted spiral-deformation feature vector $x$. The target labels are encoded as $y \in \{0, 1\}$ using a standard label encoder. The class assignments are defined such that the Healthy class is represented by 0 (negative class) and the PD class is represented by 1 (positive class).

### 4.1. Cross-validation protocol and metrics

To ensure the robustness and reproducibility of our results, we implemented a stratified 5-fold cross-validation scheme. Data shuffling was enabled with a fixed random seed to strictly preserve the Healthy-to-PD class proportions across all folds. For each model architecture, hyperparameter optimization was conducted via an exhaustive grid search, utilizing the Area Under the Receiver Operating Characteristic (ROC-AUC) curve as the primary objective function for model selection. Upon identifying the optimal configurations, we computed cross-validated predictions to assess performance. We report a comprehensive suite of metrics, including accuracy, ROC-AUC, and the confusion matrix, alongside per-class precision, recall, and F1-score (pooled across folds). Additionally, ROC curves were generated using the cross-validated predicted probabilities to visualize the diagnostic trade-offs of each classifier.

### 4.2 Model Selection.

Leveraging the proposed set of 20 spiral-based features, we conducted a systematic comparative analysis of six representative machine learning classifiers: Support Vector Machine (SVM), Logistic Regression (LR), Random Forest (RF), Gradient Boosting (GB), XGBoost (XGB), LightGBM and Vanila CNN. To ensure an unbiased performance assessment, all models were rigorously evaluated under a uniform stratified cross-validation protocol. Furthermore, to address the inherent class imbalance between Healthy and PD samples, we implemented class-weighting or minority-class upweighting strategies across all architectures. The specific hyperparameter configurations and the grid search ranges optimized for each model are summarized in Table 1.

*Table 1. hyperparameter configurations and the grid search ranges*

| **Classifier** | **Implementation Details** | **Hyperparameters & Search Space** |
|---|---|---|
| **SVM-RBF** | RBF kernel; Class-balanced weights | Regularization strength(C) ∈ {0.3, 1, 3, 10}; kernel width (γ) ∈ {0.001, 0.003, 0.01, 0.03} |
| **Log. Reg.** | ℓ2 regularization; Class-balanced | The inverse regularization strength (C-1) ∈ {0.1, 1, 10} |

| **Rand. Forest** | 400 trees; Class-balanced weights | Max depth ∈ {None, 10, 20}; Min. samples split ∈ {2, 5} |
|---|---|---|
| **Grad. Boost** | Decision tree base; Deviance loss | n_estimators ∈ {200, 400}; Learning rate (lr) ∈{0.05, 0.1} |
| **XGBoost** | Logistic objective; Ratio-based upweighting to address class imbalance | n_trees ∈ {200, 400}; Max depth ∈ {3, 6}; Learning rate ∈ {0.05, 0.1, 0.2}; Subsample ∈ {0.8, 1.0} |
| **LightGBM** | Histogram-based; Class-balanced weights | n_trees ∈ {200, 400}; Max leaves ∈ {15, 31, 63}; lr ∈ {0.05, 0.1, 0.2} |
| **Vanila CNN** | 3 convolutional layers with dropout and data augmentation | optimizer=adam, batch_size =8, loss=binary_crossentropy, input size = 128x128x3 |

## 5. Results- Observations, Validations and Analysis

### 5.1 Validation of Center Detection Strategy

To empirically validate the center-detection stage, we conducted an ablation study comparing three alternatives: foreground centroid (Centroid), distance-transform peak (DT peak), and the proposed hybrid center estimator. For each method, the full radial sampling, feature extraction, and Random Forest classification pipeline was rerun using identical 5-fold stratified cross-validation. As shown in Table 2, the proposed method achieved the highest ROC-AUC (0.751), along with the best feature extraction completeness, measured by mean angular coverage (0.791) and lowest mean maximum gap (59.24°). Although the centroid baseline yielded slightly higher classification accuracy (0.690 vs. 0.655), it produced lower coverage and substantially larger missing angular regions. These results indicate that the proposed center detector provides a stronger basis for robust spiral feature extraction and improves discrimination performance when assessed by ROC-AUC.

*Table 2. Empirical comparison of center-detection methods for spiral feature extraction and classification*

| Center method | Mean coverage fraction | Mean number of gaps | Mean max gap (deg) | RF accuracy | RF ROC-AUC |
|---|---|---|---|---|---|
| **Proposed hybrid** | 0.791 | 1.198 | 59.239 | 0.655 | 0.751 |
| **Centroid** | 0.696 | 1.399 | 116.753 | 0.690 | 0.743 |
| **DT peak** | 0.258 | 1.072 | 277.672 | 0.629 | 0.746 |

### 5.2 Sensitivity Analysis of Harmonic Band Definitions

To assess whether the frequency-domain representation was overly dependent on the manually selected harmonic bands, we performed a sensitivity analysis using several nearby alternative partitions within harmonics 2–12. Specifically, the original band definition (2–4, 5–8, 9–12) was compared against three alternatives: (2–3, 4–7, 8–12), (2–5, 6–8, 9–12), and (2–4, 5–7, 8–12). For each configuration, the feature extraction and Random Forest classification pipeline was evaluated over repeated random subsampling runs. As shown in Table 3, performance varied only modestly across the tested band definitions. The original band partition achieved the highest mean ROC-AUC (0.735), while the remaining alternatives produced comparable results. These findings indicate that the frequency-domain representation is not critically dependent on a single arbitrary band boundary and that the original partition is a reasonable and stable choice.

*Table 3. Sensitivity analysis of harmonic band definitions*

| Scheme | Low band | Mid band | High band | Mean RF Accuracy | Mean RF ROC-AUC |
|---|---|---|---|---|---|
| **Original** | 2–4 | 5–8 | 9–12 | 0.692 | 0.735 |
| **Scheme A** | 2–3 | 4–7 | 8–12 | 0.679 | 0.727 |
| **Scheme B** | 2–5 | 6–8 | 9–12 | 0.693 | 0.728 |
| **Scheme C** | 2–4 | 5–7 | 8–12 | 0.693 | 0.731 |

## 5.3 Primary Classifier Selection

Figures 4 and 5 characterize the spatial-domain feature extraction, specifically depicting the variations in radial distance from the centroid across successive revolutions for the PD samples. To complement this spatial analysis, Figures 6 and 7 provide the corresponding Fast Fourier Transform (FFT) spectra. These visualizations demonstrate that the inter-radial distance fluctuations and their associated frequency responses yield discernible patterns, providing a robust basis for differentiating between classes.

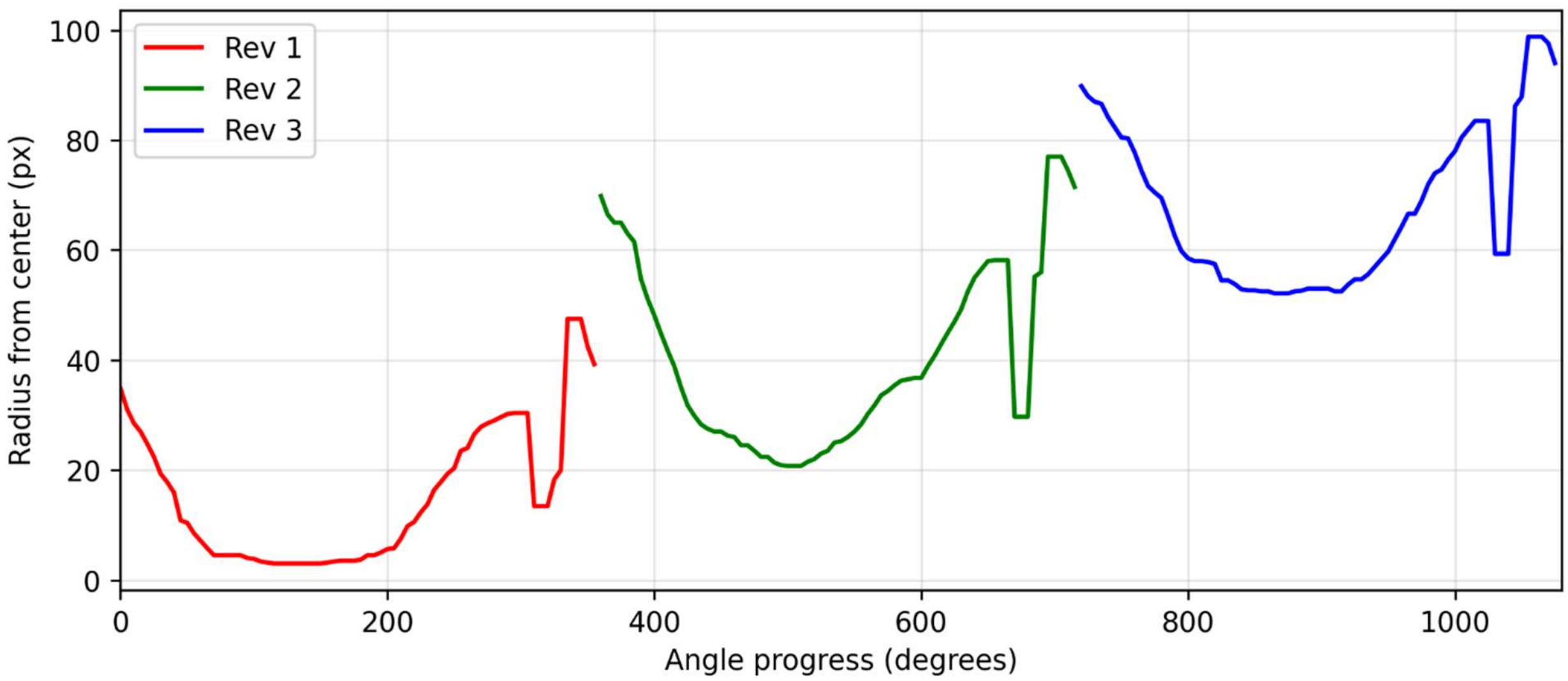

*Figure 4. Radial Distance for each Rev for Healthy sample*

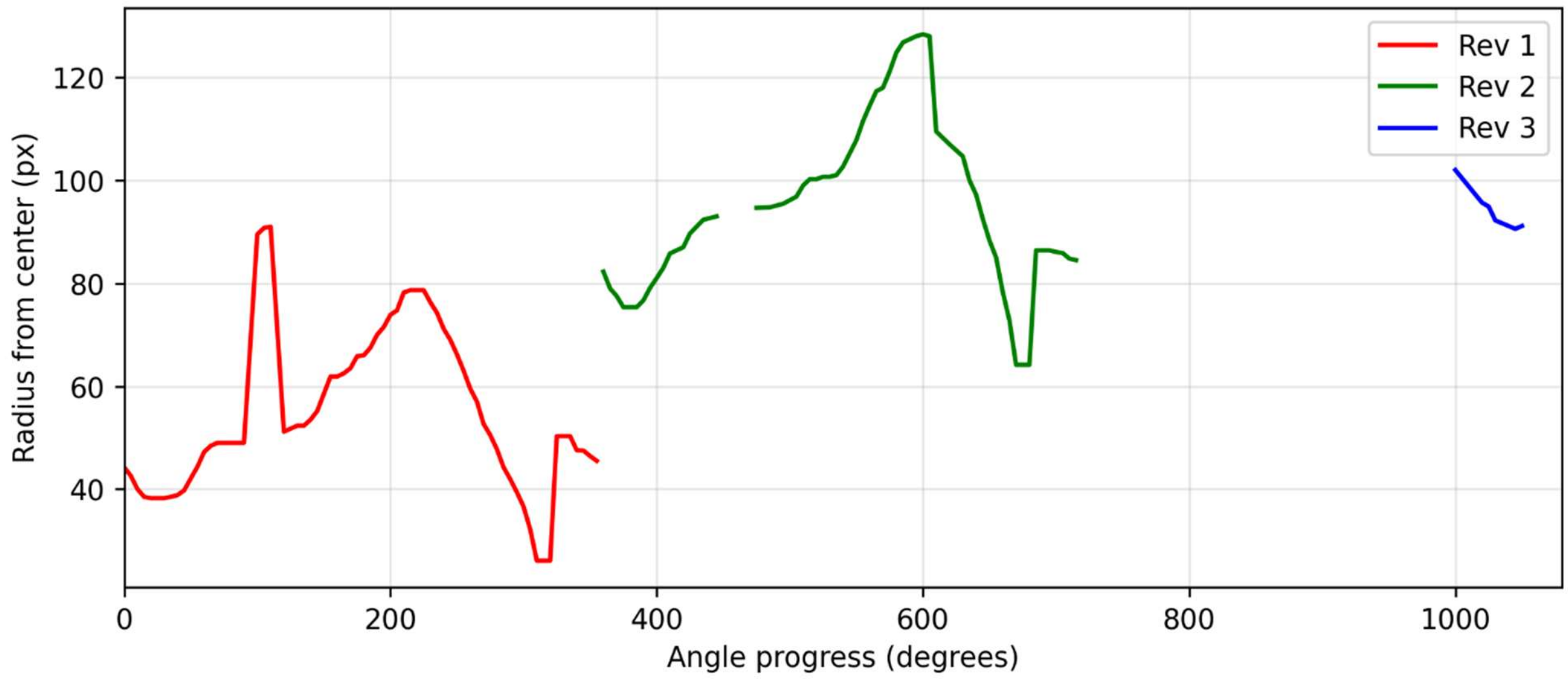

*Figure 5. Radial Distance for each Rev for PD sample*

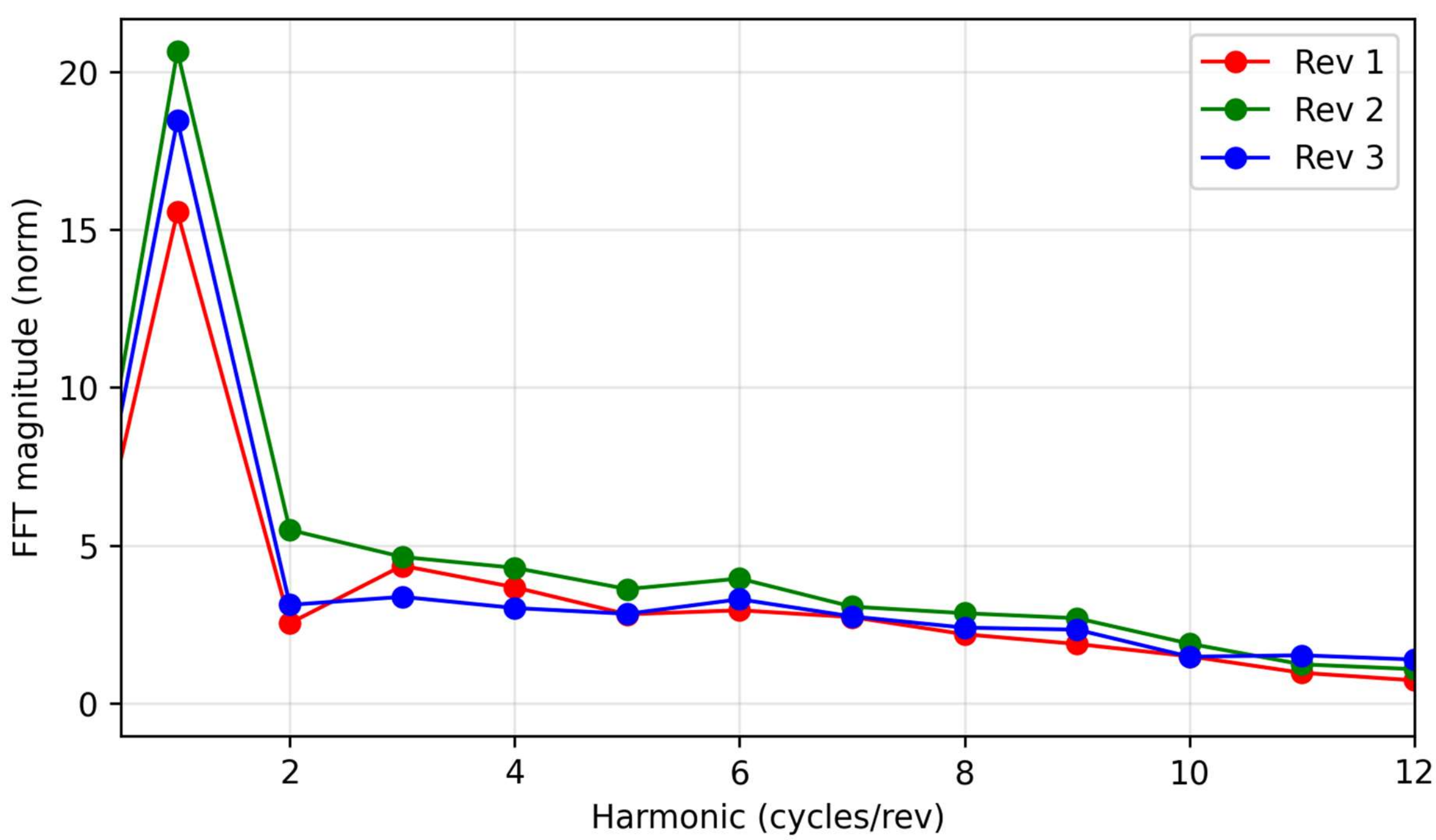

*Figure 6. FFT Magnitude plot for each Rev for Healthy*

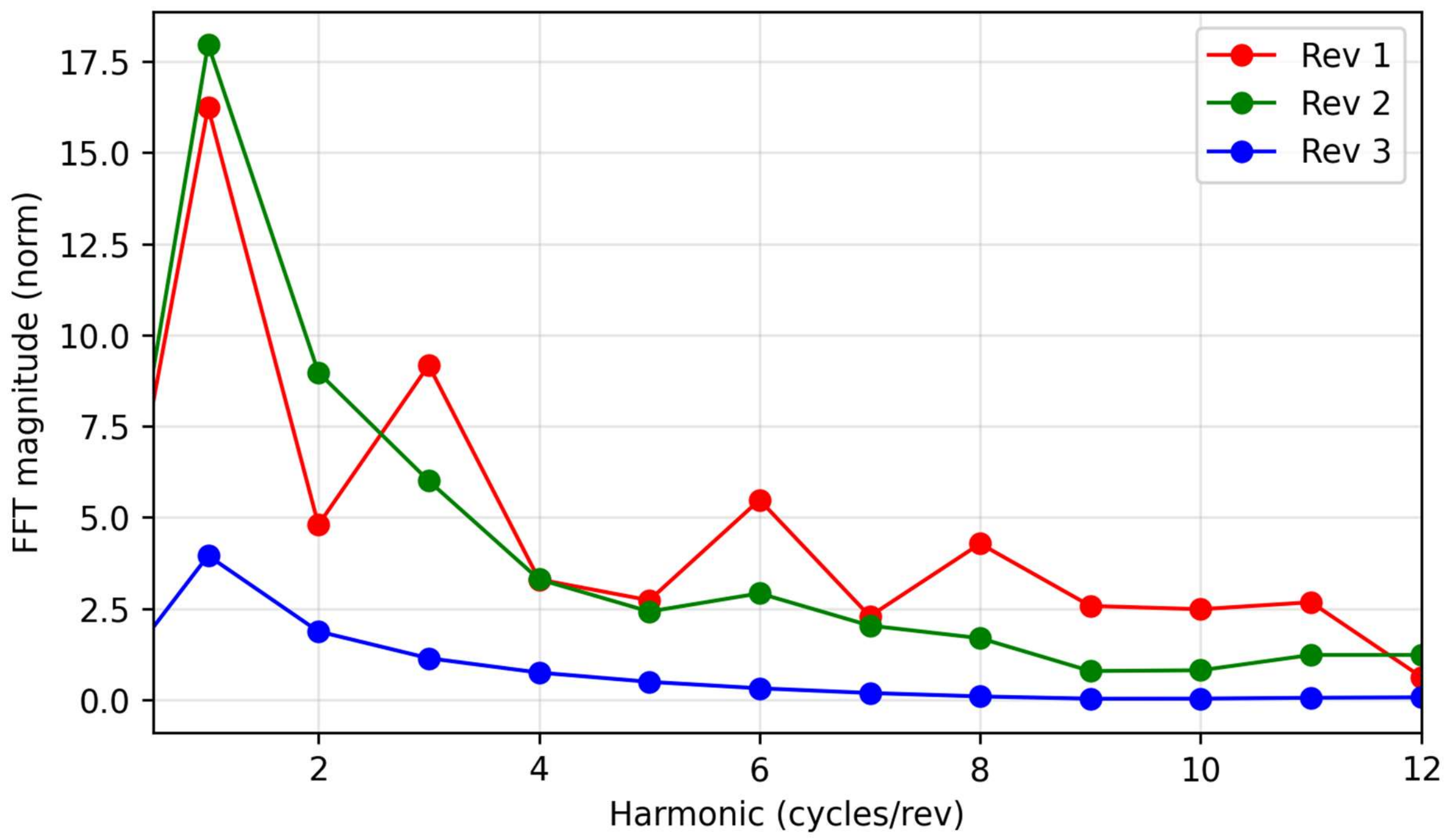

*Figure 7. FFT Magnitude plot for each Rev for PD*

The cross-validated performance and optimal hyperparameter configurations for each model are summarized in Table 4. The results indicate that the Random Forest (RF) classifier achieves the highest diagnostic performance, yielding superior results in both accuracy and ROC-AUC. The discriminative power of the RF model is further illustrated by the ROC curve in Figure 8. Consequently, based on its robust performance across all evaluation metrics, RF was selected as the primary classifier for all subsequent analyses.

*Table 4. 5-fold stratified cross-validation*

| **Model** | **Best-Hyperparamters** | **Accuracy** | **ROC–AUC** |
|---|---|---|---|
| **SVM-RBF** | C=1; γ =0.03 | 0.698 | 0.729 |
| **LR** | $C^{-1}$ =0.1 | 0.698 | 0.678 |
| **RF** | Max depth = unlim;<br>Min. samples split =2 | 0.767 | 0.832 |
| **GB** | *n_estimators* =200; lr = 0.05 | 0.724 | 0.780 |
| **XGB** | *n_trees* =200; Max depth= 6;<br>lr = 0.2; Subsample =0.8 | 0.733 | 0.821 |
| **LGBM** | *n_trees* = 400; Max leaves = 15;<br>lr= 0.2 | 0.759 | 0.831 |
| **Vanila CNN** | optimizer=adam, ConvLayers=3 loss=binary_crossentropy, metrics=accuracy, input =128x128x3 | 0.50 | 0.50 |

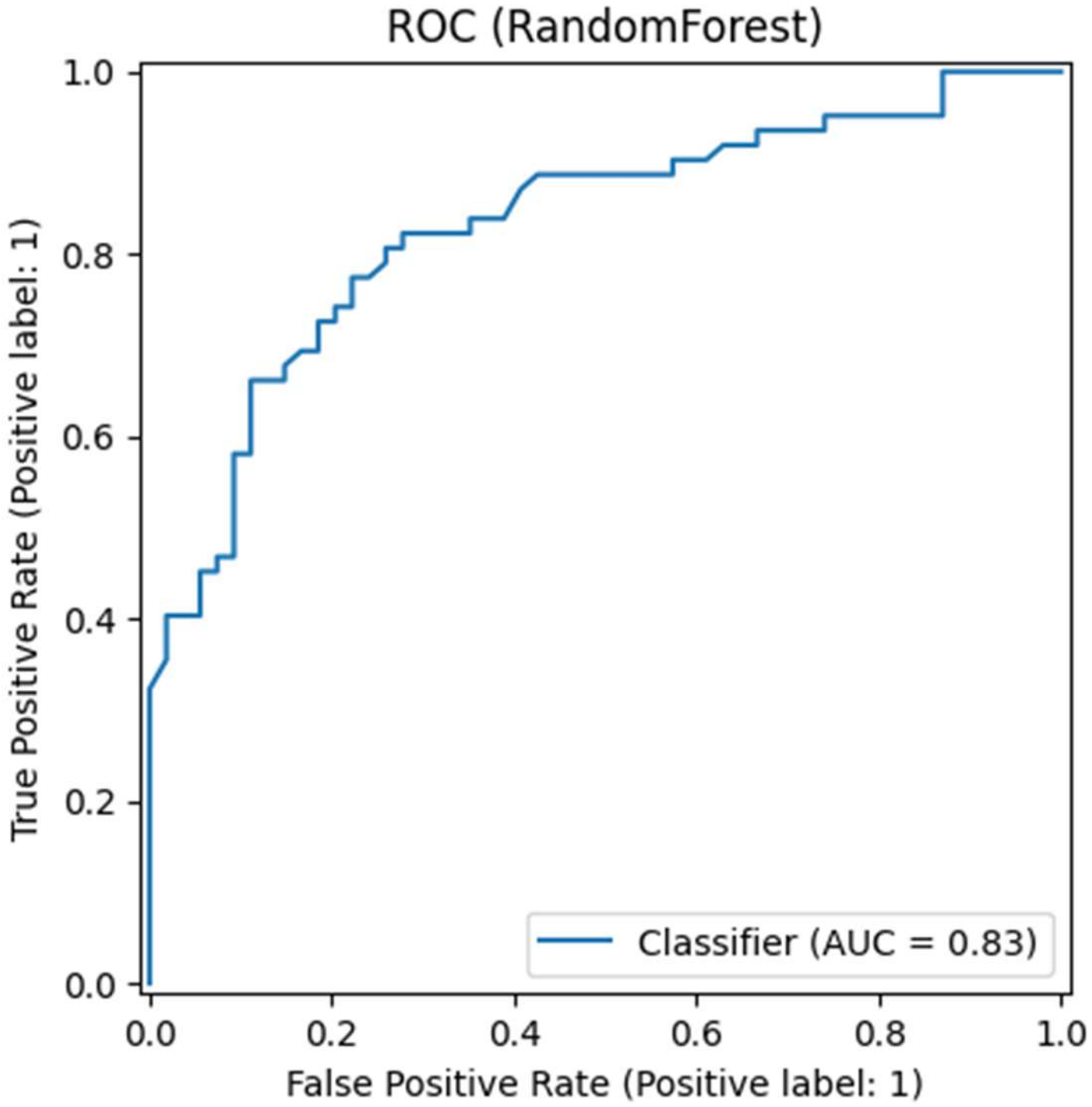

*Figure 8. ROC for the Selected Primary Classifier of RF*

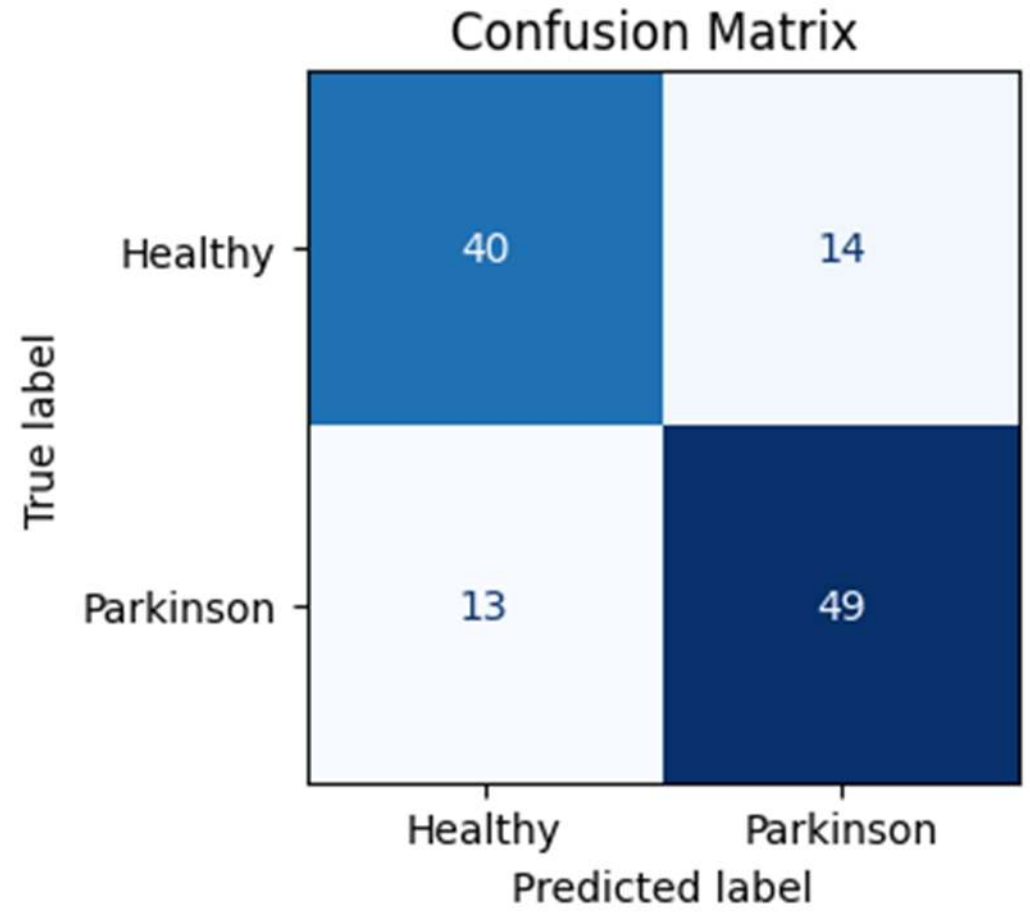

*Figure 9. Confusion Matrix for the Selected Primary Classifier of RF*

To further evaluate the diagnostic reliability of the selected Random Forest classifier, the confusion matrix is presented in Figure 9. An analysis of the classification results reveals balanced performance across both cohorts, with the model achieving an F1-score of 0.7477 for the Healthy class and 0.7840 for the PD class. This consistency across classes highlights the model's ability to minimize both false positives and false negatives, ensuring a robust diagnostic framework for biomedical application.

**5.4 Explainability: Feature Importance and Interpretation**

Given that the framework utilizes explicitly engineered radial and spectral biomarkers, we evaluate model explainability through global feature importance rankings. Figure 10 identifies the top-ranked attributes for the Random Forest classifier, highlighting the dominance of deformation-centric measures over static morphological traits.

To validate the robustness of these importance rankings, a 5-fold cross-validation stability analysis was conducted. The results of this analysis are presented in Figure 11. By calculating the Coefficient of Variation ($CV$) across 5-folds (Table 5), we observed that the primary discriminative features, specifically those related to radial derivatives and residual peak-to-peak fluctuations, exhibited high stability ($CV < 20\%$). This consistency confirms that the model is capturing reproducible pathological signals rather than over-fitting to noise within the $N = 116$ cohort. These findings reinforce the framework's utility as a stable and interpretable clinical decision-support tool.

On a different note, the model's predictive power is primarily driven by roughness and jerkiness metrics, spacing variability, and band power-derived descriptors. These results underscore a reliance on capturing the subtle high-frequency irregularities and altered stroke dynamics (micro-tremors) characteristic of Parkinsonian motor dysfunction. Unlike "black-box" deep learning architectures, the features utilized here are inherently interpretable; for instance, increased spectral energy in specific bands corresponds directly to the physical oscillations of a resting or action tremor.

To ensure clinical relevance, these importance rankings were qualitatively cross-referenced with the expert observations of the study's consulting neurologist (3rd author). The high importance assigned to frequency-domain markers provides empirical alignment with known pathological handwriting characteristics, such as diminished motor consistency and rhythmic instability. While this analysis establishes global interpretability, confirming that the model prioritizes clinically sound biomarkers, we acknowledge that "case-level" interpretability (e.g., patient-specific decision visualizations) remains an area for future development. Nonetheless, the current feature-driven approach ensures that the model's logic remains transparent and auditable by medical professionals, a key requirement for any potential clinical decision-support tool.

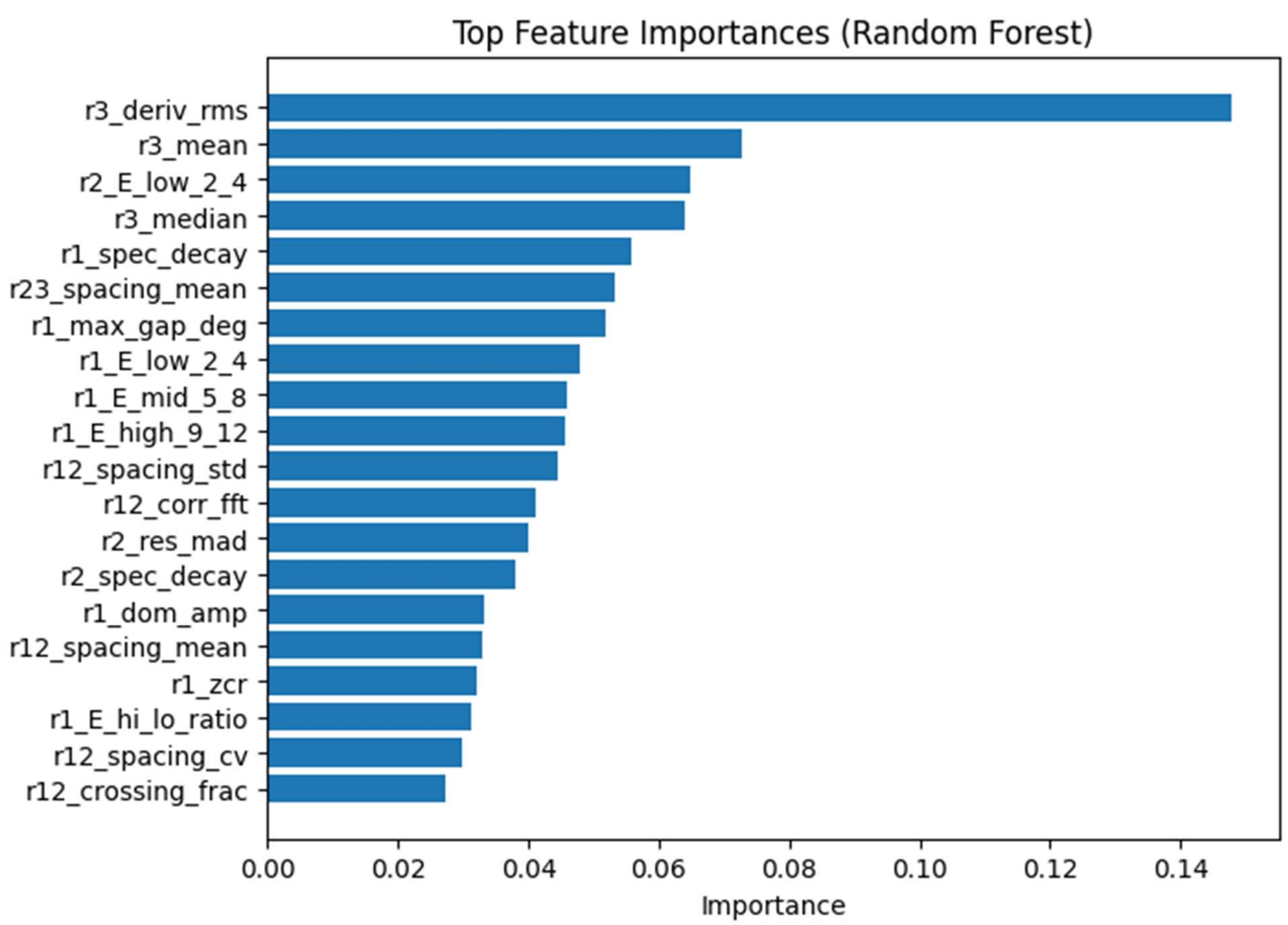

*Figure 10. Feature Importance for the Selected Classifier of RF*

*Table 5. Stability of Top 5 Discriminative Features*

| **Feature Name** | **Mean Importance** | **Std. Dev** | **Stability Rating (CV)** |
|---|---|---|---|
| **r3_deriv_rms** | 0.0496 | 0.0136 | Acceptable (27.4%) |
| **r3_dom_amp** | 0.0238 | 0.0047 | High (19.7%) |
| **r3_std** | 0.0220 | 0.0038 | High (17.5%) |
| **r1_spec_decay** | 0.0217 | 0.0042 | High (19.5%) |
| **r3_res_ptp** | 0.0194 | 0.0017 | High (8.7%) |

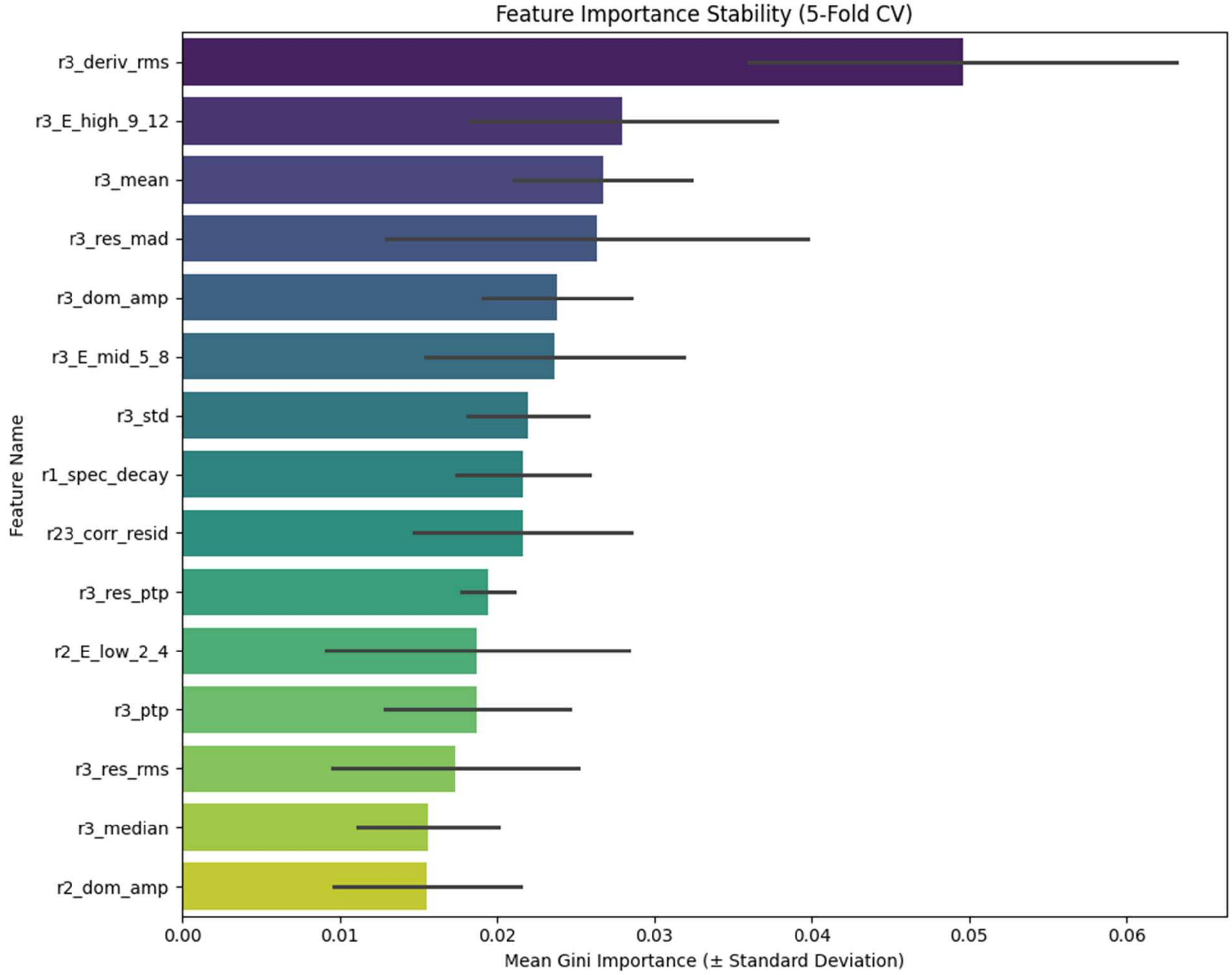

*Figure 11. Feature Importance Stability.*

## 6. Discussion

### 6.1 Clinical Significance of Radial and Spectral Biomarkers

The identification of the RMS radial derivative, $r3$_deriv_rms and $r3$_mean as the most important features for diagnosis, provides clinicians with a quantifiable measure of motor fluctuations. This aligns with prior work showing that digital spiral analysis can serve as a quantitative outcome measure for tremor severity in movement disorders[42]. In clinical practice, these metrics can serve as an automated "tremor signature." For instance, a high $r_{1_\text{hilo_ratio}}$ value directly correlates with the intensity of micro-tremors, allowing for objective monitoring of medication efficacy. Furthermore, the 3-revolution radial sampling approach ensures that the fatigue of the patient often visible in the transition from the inner to the outer revolution is captured and measured, providing insights into motor endurance that are often missed in traditional visual assessments.

### 6.2 Generalizability and Robustness

A key strength of this study is the high degree of source heterogeneity within our dataset. By integrating expert-verified clinical samples, collected under medical supervision across the Central Province of Sri Lanka, with diverse global public databases, our framework was forced to learn digitized spiral features that remain invariant across different hardware, collection environments, and demographics. This multi-source training serves as a natural stress test, mitigating the risk of cohort-specific overfitting and suggesting that the reported performance is grounded in fundamental pathological dynamics rather than institutional or device-specific artifacts.

This robustness is further validated by the necessity of our specific feature-engineering approach over automated feature learning. To verify this, a vanilla CNN was trained and evaluated on the same raw 2D image data to see if an end-to-end architecture could autonomously navigate this data heterogeneity and scarcity. However, the deep learning model failed to transcend random-chance accuracy (50.00%), illustrating that without a massive training volume, end-to-end architectures are ill-suited for this specific data scale. In contrast, by utilizing the Random Forest (RF) model alongside our targeted radial and spectral biomarkers, we successfully filtered the environmental noise that caused the CNN to fail, allowing for a highly discriminative and generalizable diagnostic performance.

### 6.3 Limitations and Future Work

While the proposed framework demonstrates high diagnostic potential, several limitations remain. First, although the total sample size (N=116) and the integration of diverse public data along with locally collected data at centralised location enhance robustness, the cohort lacks detailed longitudinal characterization. Future iterations will involve larger multi-center cohorts to confirm these findings across broader demographics.

Second, this study was designed as a binary classification between PD and healthy controls. While this established a proof-of-concept, the current model does not yet provide a differential diagnosis against "look-alike" conditions such as Essential Tremor (ET) or dystonia, which also manifest as spiral drawing abnormalities. To maintain high specificity in this pilot stage, our clinical team performed a strict differential diagnosis; however, the absence of a disease control group remains a limitation that we intend to address in future work.

Third, while our cohort consisted exclusively of treatment-naïve patients to ensure a baseline of motor symptoms unaffected by pharmacological intervention, we acknowledge that the lack of MDS-UPDRS scores and other clinical scales limits the depth of our analysis. At this stage, we have not provided a correlation analysis between the extracted computational features and established clinical severity scales. Bridging this gap is a priority for our future research to validate these digital features as true clinical biomarkers.

Finally, the current study suggests that the system is best suited as a clinical decision-support tool (CDST) rather than a standalone diagnostic platform, particularly for screening in low-resource environments. Future research will focus on expanding the dataset to include disease controls and longitudinal tracking to evaluate the model's sensitivity to disease progression and treatment efficacy.

**7. Conclusion**

This study demonstrates that the spatial and frequency-domain analysis of spiral trajectories constitutes a robust, lightweight, and computationally efficient framework for the automated detection of Parkinson's Disease. By characterizing spiral morphology through the lenses of radial fluctuations, kinematic smoothness, and turn-to-turn consistency, the proposed method achieves a significantly accurate detection with clinically relevant biomarkers without the extreme data dependency or "black-box" kind traditional deep learning architectures. Our findings indicate that features derived from the outer revolution's radial derivative ($r$3_deriv_rms), spacing variability, and high-frequency spectral energy are the most discriminative factors in distinguishing PD subjects from healthy controls. Specifically, the analysis of irregular revolution overlaps and the spectral balance of high versus low frequency content provides high diagnostic precision.

The inherent transparency of this feature-driven approach ensures that the model's outputs are directly aligned with the neurological understanding of motor dysfunction, making it an ideal candidate for integration into clinical decision-support systems where explainability is paramount. To identify the most effective classifier, we evaluated several distinct machine learning architectures, including linear (Logistic Regression), kernel-based (SVM), and ensemble methods (Gradient Boosting, XGBoost, and LightGBM). Random Forest (RF) emerged as the most robust model for this small-sample regime. Specifically, our framework with RF achieves robust performance (AUC=0.832) by training on a heterogeneous dataset that integrates expert-verified clinical samples from across the Central Province of Sri Lanka with global public data. This approach ensures feature invariance across different hardware and demographics, effectively mitigating the risks of cohort-specific overfitting and reinforcing the model's potential for real-world clinical deployment.

While the current results are promising, several avenues exist for extending this research. Longitudinal monitoring of patients could track disease progression and assess treatment response over time. Furthermore, we aim to enhance performance by fusing frequency-domain attributes with high-resolution kinematic data, such as pen velocity, acceleration, and axial pressure. Finally, conducting broader clinical trials across more diverse, large-scale datasets will be essential to ensure the generalizability and reliability required for real-world healthcare deployment.

---

**Ethics approval and consent to participate:** Ethical clearance was obtained (approval no.: THP/ PLANNING/ECR/19/2024; date: September 2, 2025) and granted by the Ethical Review Committee of the Teaching Hospital Peradeniya (University Hospital of the University of Peradeniya), Sri Lanka.

**Consent for publication:** Not applicable, as no direct clinical data are disclosed in this publication.

**Funding:** None.

**Conflict of interest:** The authors declare they have no competing interests.

**Availability of data:** The sprial image data that support the findings of this study are available on request from the corresponding author. Custom codes and scripts supporting this reserach are publicly available at https://github.com/tharaka-w/PD/ blob/main/Full_Framework_new.ipynb.